\documentstyle[twoside,epsfig,fleqn,espcrc2]{article}

\newcommand{\AmS}{{\protect\the\textfont2
  A\kern-.1667em\lower.5ex\hbox{M}\kern-.125emS}}

\title{
\hfill{\vbox{ \hbox{hep-ph/0001015}  
\hbox{FTUV/00-03} 
\hbox{IFIC/00-03}}}\\
Status of the MSW Solutions to the Solar Neutrino Problem}

\author{M. C. Gonzalez-Garcia and C. Pe\~na-Garay \address{
Instituto de F\'{\i}sica Corpuscular IFIC CSIC--Universidad de Valencia,\\
     Edificio Institutos de Paterna, Apartado 2085, 46071 Valencia}    
        \thanks{
        Talk given at the  Sixth International Workshop on
         TOPICS IN ASTROPARTICLE AND UNDERGROUND PHYSICS (TAUP99),
         September 6-10, 1999, College de France, Paris (France).
         This work was supported by Spanish DGICYT under grants 
         PB95-1077 and PB97-0693, and by the European Union TMR network 
         ERBFMRXCT960090.}}
       
\begin{document}

\maketitle
In this talk we present the results of an updated global analysis of 
two-flavor MSW solutions \cite{fitmsw} to the solar neutrino problem 
in terms of conversions of $\nu_e$ into active or sterile neutrinos including 
the the full data set corresponding to the 825-day Super--Kamiokande data 
sample as well as to Chlorine, GALLEX and SAGE experiments.

It is already three decades since the first detection of solar
neutrinos. It was realized from the very beginning that the observed
rate at the Homestake experiment was far lower than
the theoretical expectation based on the standard solar model
with the implicit assumption that neutrinos created in
the solar interior reach the earth unchanged, i.e. they are massless
and have only standard properties and interactions. 
From the experimental point of view much progress has been done in
recent years. We now have available the results of five experiments,
the original Chlorine experiment at Homestake \cite{homestake}, the
radio chemical Gallium experiments on pp neutrinos, GALLEX
and SAGE~\cite{gallium}, and the water Cherenkov 
detectors Kamiokande and Super--Kamiokande
\cite{sk700} which we summarize in Table \ref{rates}.
\begin{table*}[hbt]
\caption{Measured rates for the Chlorine, Gallium ( in SNU) 
and Super--Kamiokande (in $10^{6}$cm$^-2$s$^{-1}$) 
experiments together with the predictions from the SSM \protect{\cite{BP98}}.}
\label{rates}
\begin{tabular}{|c|c|c|c|c|c|c|c|c|c|c|}
\hline
Rates SSM & pp & pep & hep & Be & B & N & O & F &  Total & $R^{exp}$\\
\hline
Gallium &~70.45 &~2.87 &~0.015 &~34.45 &~12.38 &~3.65 &~6.04 &~0.072 
& $130\pm 7$ & $72.3 \pm 5.6$ \\
Chlorine  & ~0.0 & ~0.233 & ~0.009 & ~1.15 & ~5.89 & ~0.099 & ~.37 & ~.0044 
& $7.8 \pm 1.1$ & $2.56 \pm 0.23$\\
Super--Kam & ~0.0 & ~0.0 & ~0.00049 &~0.0 & 5.2 & ~0.0 & ~0.0 & ~0.0 
& $5.2 \pm 0.9$ & $2.45 \pm 0.23$ \\
\hline
\end{tabular}
\end{table*}
Super--Kamiokande has been able not only to confirm the original 
detection of solar neutrinos at lower rates than predicted by standard
solar models, but also to demonstrate directly that the neutrinos come
from the sun by showing that recoil electrons are scattered in the
direction along the sun-earth axis. We now have good information on
the time dependence of the event rates during the day and night, as
well as a measurement of the recoil electron energy spectrum.  After
825 days of operation, Super--Kamiokande has also presented
preliminary results on the seasonal variation of the neutrino event
rates  an issue which will become important in discriminating the MSW
scenario from the possibility of neutrino oscillations in
vacuum \cite{ourseasonal}.

In our study we use the following observables:
\begin{itemize}
\item the three measured rates shown in Table~\ref{rates} 
\item Super--Kamiokande results on the zenith angular dependence of
the event rates during 1 day and 5 night periods.
\item Recoil e spectrum including the 2 points from obtained with the
super low energy threshold below 6.5 GeV and 18 points with 6.5<E<15 MeV.
\item Seasonal variation of the event rates measured in 8 periods 
of 1.5 months each. 
\end{itemize}
We obtain the allowed value of the parameters and the corresponding
CL for the different scenarios by a $\chi^2$ analysis, details of
which can be found in Ref. \cite{fitmsw}.

In our calculations of the expected values for these observables we use as 
SSM the fluxes from Ref. \cite{BP98} but we 
also consider departures of the SSM by allowing arbitrary $^8B$ and $hep$
fluxes. For the Chlorine and Gallium 
experiments we use improved cross
sections $\sigma_{\alpha,i}(E)$ $(\alpha = e,\,x)$ from
Ref.~\cite{prod}. For the Super--Kamiokande experiment we calculate
the expected signal with the differential
cross section $d\sigma_\alpha(E_\nu,\,T')/dT'$, that we take from
\cite{CrSe} taking into account the finite energy resolution of the 
experiment which implies that the
{\em measured } kinetic energy $T$ of the scattered electron is
distributed around the {\em true } kinetic energy $T'$ according to a
resolution function which we take from~\cite{Kr97}. 
%
%

Using the predicted fluxes from the BP98 model the $\chi^2$ for the fit
to the three experimental rates is $\chi^2_{SSM}=62.4/3$dof what 
implies that the probability of the observations to be an statistical 
fluctuation of the SSM is lower than $10^{-8}$!!. One may wonder about the
possible dependence of the quality of description on the specific 
solar model used. In order to address this issue we try to fit the
data by allowing a free normalization of the dominant neutrino fluxes
$pp$ $^7Be$ and $^8B$ only imposing the constraint that the luminosity
of the sun is supplied by nuclear reactions among the light 
elements what implies a linear relation among the three normalizations.
The best fit point corresponds to an unphysical situation with negative
$^7Be$ neutrino flux. After constraining the fluxes to be positive
we obtain that the best fit point occurs at  $^7Be/^7Be_{SSM}=0$,   
$pp/pp_{SSM}=1.08$ and $^8B/^8B_{SSM}=0.53$ with 
$\chi^2_{min}=21.4/1$dof which implies that there is no acceptable fit
with a CL better than $5\times10^{-4}$\%.

Next we test the possibility of describing the data in terms of 
an energy independent neutrino conversion probability as expected,
for instance in models explaining all evidences for neutrino oscillations
(from solar, atmospheric and LSND data) in terms of three massive 
neutrinos. We find the values listed in Table \ref{pconst} 
\begin{table}[hbt]
\caption{Fit to the three measured rates for energy constant 
$\nu_e\rightarrow\nu_X$ conversion probability}
\label{pconst}
\begin{tabular}{|l|l|l|l|}
\hline
$\nu_x$ active & $P_{ee}^{best}$ & $\chi^2_{min}$/dof & CL (\%) \\
\hline
   Fix $^8B$  &  0.49 &11.7/2 &  99.71 \\
   Free $^8B$  &  0.49 &11.3/1& 99.92\\
\hline
$\nu_x$ sterile &  & & \\
\hline
   Fix $^8B$  &  0.52 &20.5/2 &99.996\\
   Free $^8B$  &  0.52 &20.5/1 & 99.99993\\
 \hline  
\end{tabular}
\end{table}
As seen in the table all scenarios with constant survival probability
are ruled out with a CL larger than 99 \%. 

\begin{figure*}[htb]
\begin{center}
\mbox{\epsfig{file=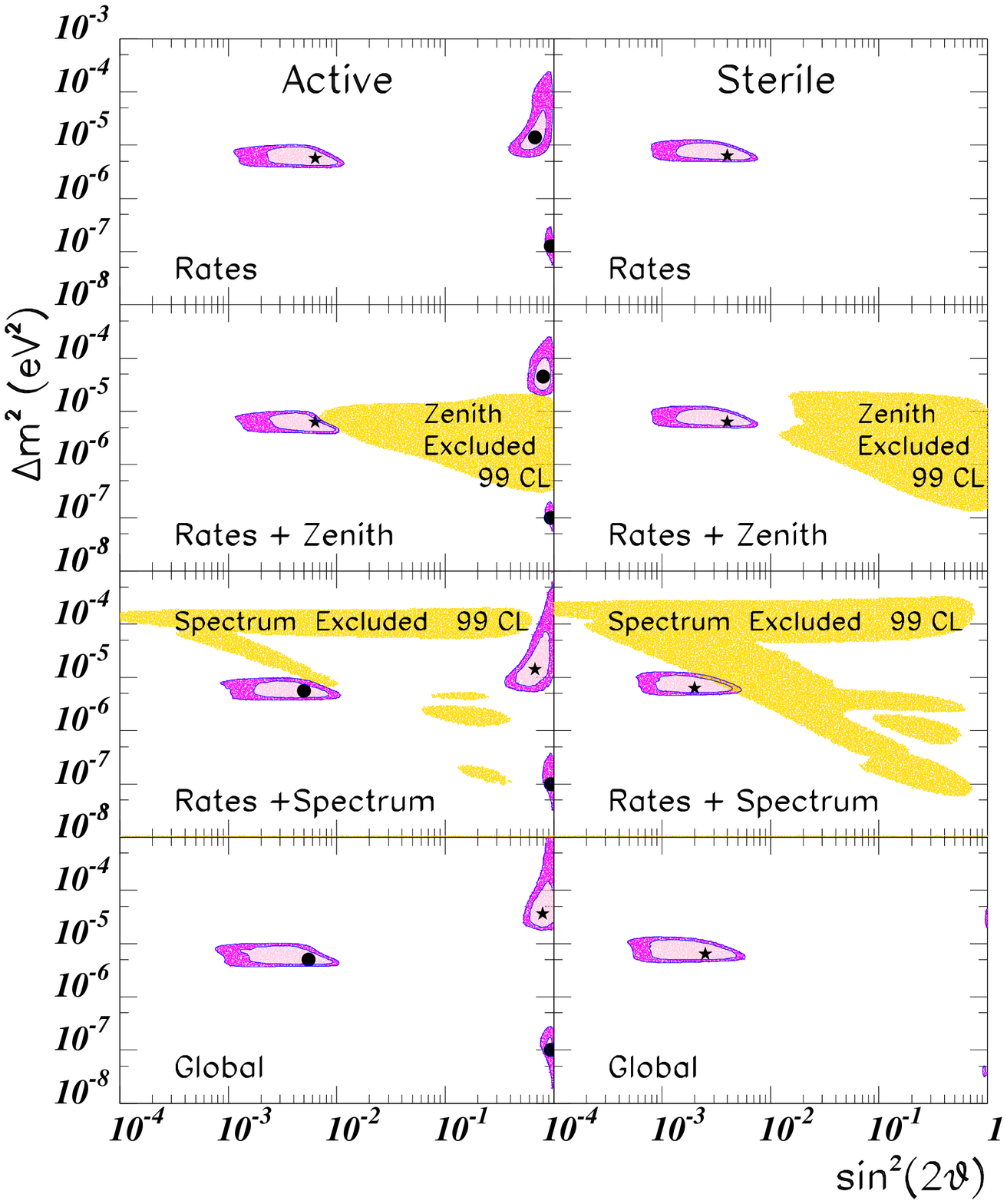,height=0.95\textheight,width=0.85\textwidth}}
\end{center}
\vglue -2cm
\caption{Allowed regions in  $\Delta m^2$ and $\sin^2\theta$ 
from the combinations of the different observables as labeled
in the figure for active-active
(left column) and active-sterile transitions (right column). 
The darker (lighter)
areas indicate 99\% (90\%)CL regions.  Global (local) best--fit points
are indicated by a star (dot).
The shadowed areas in the second (third) row represent the
region excluded by the zenith angle distribution (spectrum) data at 99\% 
CL.}
\label{fig:global}
\end{figure*}

We now consider the description in terms of matter enhanced neutrino
oscillations via the MSW mechanism \cite{msw}. We use in this case the
neutrino survival probabilities in the presence of matter given 
in Ref. \cite{LZ}. We show in Fig .\ref{fig:global} the allowed
regions for oscillations into active and sterile neutrinos 
for different combinations of the observables. In 
Table \ref{global} we give the location of the best fit point
for the different regions as well as the corresponding CL. 
\begin{table*}[hbt]
\caption{Best fit points and the corresponding AL for the different 
MSW solutions to the solar neutrino deficit for different combinations
of observables}.
\label{global}
\begin{tabular}{|c|c|c|c|c|c|}
\hline
& & rates & rates+zenith  & rates+spectrum & rates+zenith+spectrum+season \\
\hline
MSW& dof& 1 & 6 & 18 & 30 \\
\hline
& $\Delta m^2$ & $5.6\times 10^{-6}$ & $5.0\times 10^{-6}$ &
$5.6\times 10^{-6}$ & $5.1\times 10^{-6}$ \\
SMA & $\sin^2(2\theta)$ & 0.0063 & 0.0063 & 0.005 & 0.0055 \\
& $\chi^2_{min}$ (\%CL)  
& {\bf 0.37} (55)   & {\bf 5.9} (56)  & 23.4 (83) &  37.4 (83)\\
\hline  
& 
$\Delta m^2$ & $1.4 \times 10^{-5}$   & $4.5 \times 10^{-5}$    
& $1.4 \times 10^{-5}$  & $3.6 \times 10^{-5}$    \\
LMA & $\sin^2(2\theta)$ & 0.67   &  0.8 & 0.67  & 0.79  \\
& $\chi^2_{min}$  (\%CL)  & 2.92 (91)  & 7.2 (70)& {\bf 22.5} (79) 
&  {\bf 35.4} (77)\\
\hline  
& $\Delta m^2$ &  $1.3 \times 10^{-7}$ & $1.0 \times 10^{-7}$ &
$1.0 \times 10^{-7}$ & $1.0 \times 10^{-7}$ \\
LOW & $\sin^2(2\theta)$ & 0.94   & 0.94  & 0.94  &  0.94 \\
& $\chi^2_{min}$ (\%CL)   &7.4 (99)  & 12.7 (95) & 26.7 (91.5) &  40. (90)\\
\hline
& $\Delta m^2$ & $5.\times 10^{-6}$ & $5.0\times 10^{-6}$ &
$5.\times 10^{-6}$ & $5.\times 10^{-6}$ \\
SMA & $\sin^2(2\theta)$ & 0.005 & 0.005 & 0.003 & 0.003 \\
sterile & $\chi^2_{min}$ (\%CL) &2.6 (90)  &8.1 (77)   & 26.3 (89) & 40. (90) 
\\
\hline    
\end{tabular}
\end{table*}
Several comments are in order. First we see that 
for active neutrinos there are three allowed regions, the 
small mixing angle region (SMA), the large mixing angle
region (LMA) and the low mass region (LOW). 
For sterile neutrinos only the SMA solution is allowed.
This arises from the fact that 
unlike active neutrinos which lead to events in the
Super--Kamiokande detector by interacting via neutral current with the
electrons, sterile neutrinos do not contribute to the Super--Kamiokande
event rates.  Therefore a larger survival probability for $^8B$
neutrinos is needed to accommodate the measured rate. As a consequence
a larger contribution from $^8B$ neutrinos to the Chlorine and Gallium
experiments is expected, so that the small measured rate in Chlorine
can only be accommodated if no $^7Be$ neutrinos are present in the
flux. This is only possible in the SMA solution region, since in the
LMA and LOW regions the suppression of $^7Be$ neutrinos is not enough.

As for the quality of the different solutions for the oscillations into
active neutrinos we find that the different observables favour different
solutions. The total rates favours the SMA solution while the inclusion
of the zenith angular dependence favours the LMA solution as although
small, some effect is observed in the zenith angle dependence which
points towards a larger event rate during the night than during the
day, and that this difference is constant during the night as expected
for the LMA solution. In the SMA solution, however, the
enhancement is expected to occur mainly in the fifth night. 
The spectral information is such that the oscillation hypothesis 
does not improve considerably the fit to the energy spectrum as 
compared to the no-oscillation hypothesis as the data is basically consistent
with a flat distribution what is also in better agreement with the LMA and 
LOW solutions while the SMA predicts a continuously raising spectrum.
Moreover, the observation of a possible seasonal variation of the 
higher energy event rates in Super--Kamiokande could also be accommodated 
in terms of  the LMA solution to the solar neutrino problem 
\cite{ourseasonal}.  
So once all the observables are combined we find that all solutions
are allowed at the 90 \% CL. LMA and SMA give similar descriptions,  
the LMA being slightly favoured. 

\end{document}